\documentstyle[aps,pre,12pt]{revtex}

\newcommand{\BE}{\begin{equation}}
\newcommand{\EE}{\end{equation}}
\newcommand{\BA}{\begin{eqnarray}}
\newcommand{\EA}{\end{eqnarray}}

\begin{document}
\draft

\title{Acoustic Band Gap and Coherent Behavior in a Periodic 
Two-Dimensional System}

\maketitle

\vspace{ 24 pt}

\begin{center}

{\bf Z. Ye$^\star$, E. Hoskinson$^{\star,\dagger}$, and  H. Hsu$^\star$}

\end{center}

\vfill

\begin{center}

{\bf April 24, 1999}

\end{center}

\vfill

\begin{center}

{
$^\star$Department of Physics, National Central University, Chung-li,
Taiwan 320, ROC\\ $^\dagger$Department of Physics, University of California,
Berkeley, CA 94720, USA}

\end{center}

\newpage





Many important discoveries in physics this century originate from 
the research of wave propagation in periodic structures\cite{Russell}.
An exciting example is the discovery of what is called {\it wave band 
structures}\cite{John}, asserted by Bloch theorem. 
Of special importance is the appearance of
the complete band gaps in strong scattering structures. In
the gaps, the system has no propagation modes and waves decay 
exponentially in any direction, 
rendering applications such as in noise screening and 
laser devices.
Band structures or complete band gaps have been observed or
predicted for electromagnetic waves in periodic dielectric 
structures\cite{Ho,Yab}, and only recently for acoustic 
waves\cite{Kushwaha,Weaver,Sanchez,Torres} and hydrodynamic 
waves\cite{TorresNature}. 

We have considered the acoustic band structures in a periodic two-dimensional
system. Here we show that not only a complete band gap may be achieved, but
an interesting collective behavior may appear in such a system. 
Place uniform air-cylinders of radius $a$ in parallel to form a square
lattice in water perpendicular to the $x-y$ plane; thus the cylinders are
parallel to
the $z$-axis in the setting. The fraction of area occupied by the cylinders
per unit area is $\beta$; thus the lattice constant $d$ is
$(\pi/\beta)^{1/2}a$.
Due to the large contrast in acoustic impedance between water and air, the
air-cylinders are strong acoustic scatterer, and because of this the
complete gap is
relatively easy to realize. The scattering function of one of such
cylinders can
be readily computed in the form of a modal series solution, having a
resonance scattering peak near $ka = 0.005$.

The band structure in the system is presented in Fig.~1(a), for which the 
cylinder concentration $\beta$ is taken as $10^{-3}$ and the
frequency is written in terms of the non-dimensional parameter $ka$. A
complete band gap appears between around $ka = 0.007$ and $0.032$, somewhat
above the resonance peak of a single cylinder. The complete gap is found to
be stable against small distortions of the lattice structure.
The width of the gap, however, diminishes as the cylinder concentration
$\beta$ decreases. Fig.~1(b), showing the dependence of the width on $\beta$,
indicates that the gap disappears when $\beta$ drops below $10^{-5}$. While
the low band edge does not change much as $\beta$ varies, the higher edge
seems to increase exponentially with $\beta$ in the range of frequencies
considered.

To investigate what happens to the cylinders when the complete gap 
occurs, a novel diagram method is used to characterize the behavior of the 
cylinders themselves. Suppose that a unit line source, transmitting 
monochromatic waves, is located at the center of the array of the cylinders 
and aligned parallel to the cylinders. The transmitted waves are multiply
scattered by the cylinders, causing the scattering
characteristics of the scatterers to change. 
The scattered wave from each cylinder is a
response to all incident waves consisting of the direct wave from the source
and the scattered waves from the other cylinders. The total wave, a measure
of the transmitted energy, will be the summation of the direct wave and all
the scattered waves at any spatial point. Driving by the source and
the scattered waves, the surface of each cylinder vibrates.
Invoking usual boundary conditions at the air-water interface, the
surface vibration can be computed rigorously
by the self-consistent method\cite{Foldy} and the matrix inversion\cite{PRL}.

Express the surface vibration for a cylinder, the $i$-th cylinder say,
as $|A_i|e^{j\theta_i}$, with $j = \sqrt{-1}$ and $i = 1, 2, \dots$.
In the range of frequencies concerned, the pulsating mode of vibration
dominates. The modulus $|A_i|$ represents the vibration magnitude, and
$\theta_i$ is the phase. Assign a unit vector $\vec{u}_i$, hereafter termed
phase vector, to associate with each phase $\theta_i$ such that
$\vec{u}_i = \cos\theta_i \vec{e}_x + \sin\theta_i \vec{e}_y$.
Setting the phase of the source to zero, numerical experiments
are carried out to study the behavior of the phase vectors of the
cylinders and the spatial distribution of the acoustic energy.

Three transmitted frequencies are chosen to be below, inside and above the
complete band gap respectively. The results are presented in Fig.~2. We
observe that for frequencies outside the gap, there is no ordering for the
phase vectors and the energy distribution is extended. The more
variations in the case of $ka=0.006$ than that of $ka=0.034$ imply the
stronger multiple scattering in the system.
Inside the gap,
an ordering in the phase vectors appears, that is, all cylinders are
oscillating completely in phase, but exactly out of phase with the
transmitting source. Meanwhile, the energy is trapped near the
source and rapidly decays in all
directions, in accordance with the evanescent mode of the cylinder
lattice. The results indicate that the collective behavior
of the cylinders are associated with the complete band gap. Such
out-of-phase coherence allows for effective cancellation of wave
propagation.

\newpage

\section*{Figure Captions}

\begin{description}

\item[Figure 1]
(a) Acoustic bands for a square lattice of air-cylinders for
$\beta = 10^{-3}$. A complete band gap
lies between the two horizontal dashed lines.
(b) The dependence of the complete band gap on
the cylinder concentration $\beta$.

\item[Figure 2] The left column: The phase diagram for the two dimensional
phase vectors defined in the text.
Right column: The spatial distribution of acoustic energy. For brevity,
200 cylinders are considered in the computation. In the plot of
the energy distribution, the geometrical spreading factor has been removed.
Note that in the case of $ka = 0.01$
there are a few phase vectors which do not point to the same direction as the
others because of the effect of the finite boundary; when
the cylinder number goes to infinity, this effect disappears.

\end{description}

\vspace{24 pt}

\noindent {\bf Acknowledgments.}
The work received support from the National
Central University and the National Science Council.

\vspace{24pt}

\noindent
Correspondence and request for materials should be addressed to
Z.Y. \\
Email : zhen@joule.phy.ncu.edu.tw


\begin{references}



\bibitem{Russell} Russell, P. St. J. Photonic band gaps. Physics World,
August, 37-42 (1992).

\bibitem{John} John, S. Localization of light. Physics Today {\bf 44}, 
32-40 (1991). 

\bibitem{Ho} Ho, K. M., Chan, C. T., and Soukoulis, C. M. Existence of
a photonic band gap in periodic dielectric structures. Phys. Rev. Lett.
{\bf 65}, 3152-3155 (1990).

\bibitem{Yab} Yablonovitch, E. and Gmitter, T. J. Photonic band structure: 
The face-centered-cubic case employing nonspherical atoms.
Phys. Rev. Lett. {\bf 67}, 2295-2298 (1991).

\bibitem{Kushwaha} Kushwaha, M. S., Halevi, P., Dobrzynski, L.,
and Djafari-Rouhani, B. Acoustic band structure of periodic elastic
composites. Phys. Rev. Lett. {\bf 71}, 2022-2025 (1993). 

\bibitem{Weaver} Turner, J. A., Chambers, M. E., 
and Weaver, R. L. Ultrasonic band gaps in aggregates of sintered aluminum
beads. Acustica, {\bf 84}, 1-4 (1998).

\bibitem{Sanchez} Martinez-Sala, R. et al. Sound attenuation by sculpture.
Nature {\bf 378}, 241 (1995).

\bibitem{Torres} Montero de Espinosa, F. R., Jimenez, E., and
Torres, M. Ultrasonic band gap in a period two-dimensional composite.
Phys. Rev. Lett. {\bf 80}, 1208 (1998).

\bibitem{TorresNature} Torres, M., Adrados, J. P., and
Montero de Espinosa, F. R. Visualization of Bloch waves and domain walls.
Nature {\bf 398}, 114-115 (1998).

\bibitem{Foldy} Foldy, L. L. The multiple scattering of waves. Phys.
Rev. {\bf 67}, 107-119 (1945).

\bibitem{PRL} Ye, Z. and Alvarez, A. Acoustic localization in bubbly
liquids. Phys. Rev. {\bf 80}, 3503-3506 (1998).

\end{references}
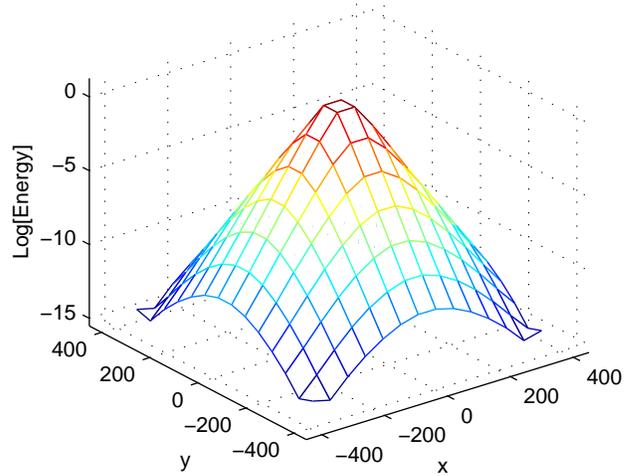
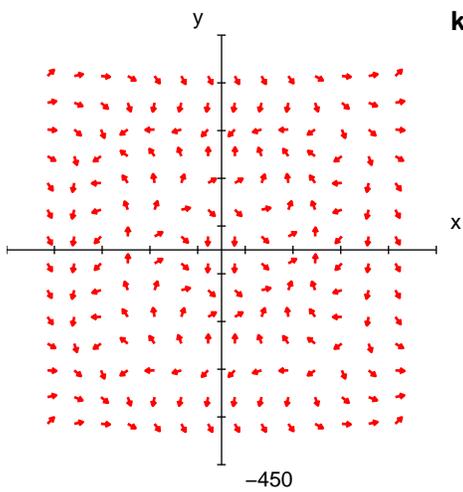
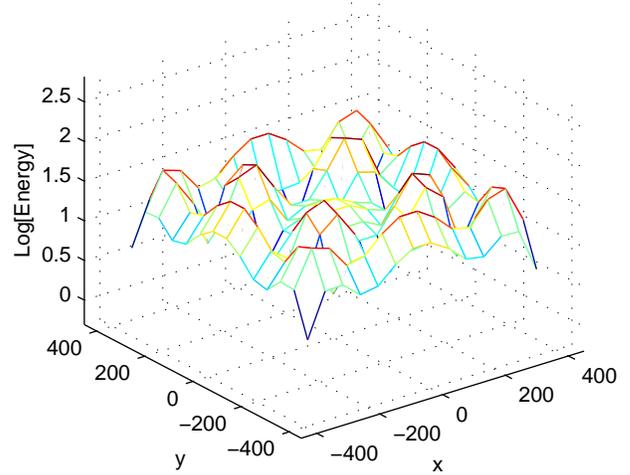
\end{document}